\documentclass[aps,prc,twocolumn,floatfix,nofootinbib,showpacs,superscriptaddress]{revtex4-1}

\usepackage{amsmath,amsfonts,amssymb,bm}
\usepackage{graphicx}
\usepackage{color}
\definecolor{purple}{rgb}{0.5,0,0.5}
\definecolor{blue}{rgb}{0.0,0,0.9}

\begin{document}

\title{Expanding the concept of in-hadron condensates}

\author{Lei Chang}
\affiliation{Physics Division, Argonne National Laboratory, Argonne, Illinois 60439, USA}

\author{Craig D.~Roberts}
\affiliation{Physics Division, Argonne National Laboratory, Argonne, Illinois 60439, USA}
\affiliation{Department of Physics, Center for High Energy Physics and State Key Laboratory of Nuclear Physics and Technology, Peking University, Beijing 100871, China}
\affiliation{Department of Physics, Illinois Institute of Technology, Chicago, Illinois 60616-3793, USA}

\author{Peter C.~Tandy}
\affiliation{Center for Nuclear Research, Department of
Physics, Kent State University, Kent OH 44242, USA}


\begin{abstract}
The in-pseudoscalar-meson condensate can be represented through the pseudoscalar-meson's scalar form factor at zero momentum transfer.  With the aid of a mass formula for scalar mesons, revealed herein, the analogue is shown to be true for in-scalar-meson condensates.  The concept is readily extended to all hadrons so that, via the zero momentum transfer value of any hadron's scalar form factor, one can readily extract the value for a quark condensate in that hadron which is a measure of dynamical chiral symmetry breaking.
\end{abstract}


\maketitle

%
Dynamical chiral symmetry breaking (DCSB) and its connection with the generation of hadron masses was first considered in Ref.\,\cite{Nambu:1961tp}.  The effect was represented as a vacuum phenomenon.  Two essentially inequivalent classes of ground-state were identified in the mean-field treatment of a meson-nucleon field theory: symmetry preserving (Wigner phase); and symmetry breaking (Nambu phase).  Notably, within the symmetry breaking class, each of an uncountable infinity of distinct configurations is related to every other by a chiral rotation.  This is arguably the origin of the concept that strongly-interacting quantum field theories possess a nontrivial vacuum.

With the introduction of the parton model for the description of deep inelastic scattering (DIS), this notion was challenged via an argument \cite{Casher:1974xd} that DCSB can be realised as an intrinsic property of hadrons, instead of via a nontrivial vacuum exterior to the observable degrees of freedom.  This perspective is tenable because the essential ingredient required for dynamical symmetry breaking in a composite system is the existence of a divergent number of constituents and DIS provided evidence for the existence within every hadron of a divergent sea of low-momentum partons.  This view has, however, received scant attention.  On the contrary, the introduction of QCD sum rules \cite{Shifman:1978bx} as a method to estimate nonperturbative strong-interaction matrix elements entrenched the belief that the QCD vacuum is characterised by numerous, independent, non-vanishing condensates.

Notwithstanding the prevalence of this belief, it does lead to problems; e.g., entailing a cosmological constant that is $10^{46}$-times greater than that which is observed \cite{Turner:2001yu,Brodsky:2009zd}.  This unwelcome consequence is partly responsible for reconsideration of the possibility that the so-called vacuum condensates are in fact an intrinsic property of hadrons.  Namely, in a confining theory, condensates are not constant, physical mass-scales that fill all spacetime; instead, they are merely mass-dimensioned parameters that serve a practical purpose in some theoretical truncation schemes but otherwise do not have an existence independent of hadrons \cite{Brodsky:2009zd,Brodsky:2008be,Brodsky:2010xf,Glazek:2011vg}.  Regarding the quark condensate, this perspective was recently elucidated for light pseudoscalar mesons \cite{Brodsky:2010xf}.  Herein we propose an extension of the concept to all hadrons.

We start with Ref.\,\cite{GellMann:1968rz}, which presents the relation
\begin{equation}
\label{gmor}
m_\pi^2 = \lim_{P^\prime \to P \to 0} \langle \pi(P^\prime) | {\cal H}_{\chi{\rm sb}}|\pi(P)\rangle\,,
\end{equation}
where $m_\pi$ is the pion's mass and ${\cal H}_{\chi{\rm sb}}$ is that part of the hadronic Hamiltonian density which explicitly breaks chiral symmetry.  It is important to observe that the operator expectation value in Eq.\,(\ref{gmor}) is evaluated between pion states.  In terms of QCD quantities, Eq.\,(\ref{gmor}) entails
\begin{eqnarray}
\label{gmor1}
\lefteqn{
\forall m_{ud} \sim 0\,,\;  m_{\pi^\pm}^2 =  m_{ud}^\zeta \, {\cal S}_\pi^\zeta(0)\,,}\rule{7.2em}{0ex}\\
{\cal S}_\pi^\zeta(0) & = & - \langle \pi(P) | \mbox{\small $\frac{1}{2}$}(\bar u u + \bar d d) |\pi(P)\rangle\,,
\label{gmor1a}
\end{eqnarray}
where $m_{ud}^\zeta = m_u^\zeta+m_d^\zeta$, $m_{u,d}^\zeta$ are the current-quark masses at a renormalisation scale $\zeta$, and ${\cal S}^\zeta(0)$ is the pion's scalar form factor at zero momentum transfer, $Q^2=0$.  The right-hand-side (rhs) of Eq.\,(\ref{gmor1}) is proportional to the pion $\sigma$-term (see, e.g., Ref.\,\cite{Flambaum:2005kc}).  Consequently, using the connection between the $\sigma$-term and the Feynman-Hellmann theorem, Eq.\,(\ref{gmor}) is actually the statement
\begin{equation}
\label{pionmass2}
\forall m_{ud} \sim 0\,,\; m_\pi^2 = m_{ud}^\zeta \frac{\partial }{\partial m^\zeta_{ud}} m_\pi^2.
\end{equation}

Recall now that one may use the axial-vector Ward-Takahashi identity to prove \cite{Maris:1997hd}: for any pseudoscalar meson, $P$, constituted from quarks $q$ and $Q$, whether ground-state, excited-state or hybrid,
\begin{equation}
\label{gmorR}
f_P m_P^2 = (m_q^\zeta + m_Q^\zeta) \rho_P^\zeta ,
\end{equation}
where $m_{q,Q}$ are the current-quark masses and
\begin{eqnarray}
\label{fpigen}
\lefteqn{i f_P K_\mu = \langle 0 | \bar Q \gamma_5 \gamma_\mu q |P \rangle} \\
\nonumber
& = & Z_2(\zeta,\Lambda)\; {\rm tr}_{\rm CD}
\int_k^\Lambda i\gamma_5\gamma_\mu S_q(k_+) \Gamma_P(k;K) S_Q(k_-)\,, 
\\
\nonumber
\lefteqn{i\rho_P^\zeta = -\langle 0 | \bar Q i\gamma_5 q |P \rangle} \\
& = & Z_4(\zeta,\Lambda)\; {\rm tr}_{\rm CD}
\int_k^\Lambda \gamma_5 S_q(k_+) \Gamma_P(k;K) S_Q(k_-) \,.\label{rhogen}
\end{eqnarray}
($K^2=-m_P^2$; $k_\pm = k\pm K/2$, without loss of generality in a Poincar\'e covariant approach.)  Here, $f_P$ is the pseudoscalar meson's leptonic decay constant and the rhs of Eq.\,(\ref{fpigen}) expresses the axial-vector projection of the $P$-meson's Bethe-Salpeter wavefunction onto the origin in configuration space.  Likewise, Eq.\,(\ref{rhogen}) describes the pseudoscalar projection of the $P$-meson's Bethe-Salpeter wavefunction onto the origin.  It is therefore just another type of $P$-meson decay constant.  Plainly then, both $f_P$ and $\rho_P^\zeta$ are intrinsic properties of the hadron.  Moreover,
\begin{equation}
\label{inpiqbq}
\kappa_P^\zeta \equiv -\langle \bar Q q \rangle^\zeta_P := -f_P \langle 0 | \bar Q \gamma_5 q |P \rangle
= f_P \rho_P^\zeta
\end{equation}
is the in-hadron condensate introduced in Ref.\,\cite{Maris:1997tm}.

We note that $\int_k^\Lambda:=\int^\Lambda \!\! \mbox{\footnotesize $\displaystyle\frac{d^4 k}{(2\pi)^4}$}$ in Eqs.\,(\ref{fpigen}), (\ref{rhogen}) represents a Poincar\'e-invariant regularization of the integral, with $\Lambda$ the ultraviolet regularization mass-scale;
$\Gamma_{P}(k;P) $ is the pseudoscalar meson's canonically-normalised Bethe-Salpeter amplitude; viz.,
\begin{eqnarray}
\nonumber
\lefteqn{\Gamma_{P}(k;K) = \gamma_5 \left[ i E_P(k;K) + \gamma\cdot K F_P(k;K) \right.}\\
&& \left. + \,\gamma\cdot k \, G_P(k;K) - \sigma_{\mu\nu} k_\mu K_\nu H_P(k;K) \right];
\label{genGpi}
\end{eqnarray}
$S_q$, $S_Q$ are the dressed-propagators of the $q$- and $Q$-quarks; and $Z_{2,4}(\zeta,\Lambda)$ are, respectively, the quark wavefunction and Lagrangian mass renormalisation constants.
%

Using Eq.\,(\ref{gmorR}), one obtains
\begin{equation}
\label{gmor2}
{\cal S}_\pi^\zeta(0)
= \frac{\partial }{\partial m^\zeta_{ud}} m_\pi^2
=\frac{\partial }{\partial m^\zeta_{ud}} \left[ m_{ud}^\zeta\frac{\rho_\pi^\zeta}{f_\pi}\right].
\end{equation}
Equation~(\ref{gmor2}) is valid for any values of $m_{u,d}$, including the neighbourhood of the chiral limit, wherein
\begin{equation}
\label{gmor3}
\frac{\partial }{\partial m^\zeta_{ud}} \left[ m_{ud}^\zeta\frac{\rho_\pi^\zeta}{f_\pi} \right]_{m_{ud} = 0}
= \frac{\rho_\pi^{\zeta 0}}{f_\pi^0} =: B_\pi^{\zeta 0}.
\end{equation}
The superscript ``0'' indicates that the quantity is computed in the chiral limit.  It is well known that $f_\pi^0 \neq 0$ if (and only if) chiral symmetry is dynamically broken in QCD: it is an order parameter for DCSB.  Less widely appreciated is that in the chiral limit the numerator is another well-known quantity; viz., using QCD's quark-level Goldberger-Treiman relations, one can prove \cite{Maris:1997hd}:
\begin{equation}
\label{gmor4}
f_\pi^0 \, \rho_\pi^{\zeta 0} 
= - \langle \bar q q \rangle^{\zeta 0}\,,
\end{equation}
where the rhs is the so-called vacuum quark condensate.  Thus, as demonstrated previously \cite{Brodsky:2010xf,Maris:1997hd,Maris:1997tm}, the vacuum quark condensate is actually the chiral-limit value of the in-pion condensate; i.e., it describes a property of the chiral-limit pion.  Importantly, Ref.\,\cite{Langfeld:2003ye} establishes that the rhs of Eq.\,(\ref{gmor4}) is precisely the same condensate that appears: as a constant in the operator product expansion \cite{Lane:1974he}; via the Banks-Casher formula \cite{Banks:1979yr}; and through the trace of the chiral-limit dressed-quark propagator.

With Eqs.\,(\ref{gmor1}), (\ref{gmor2}), (\ref{gmor3}), (\ref{gmor4}), one has shown that in the neighbourhood of the chiral limit
\begin{equation}
m_{\pi^\pm}^2 =  -m_{ud}^\zeta  \frac{\langle \bar q q \rangle^{\zeta 0}}{(f_\pi^0)^2} + {\rm O}(m_{ud}^2).
\end{equation}
Neither PCAC nor soft-pion theorems were employed in analysing the rhs of Eq.\,(\ref{gmor1}).  The analysis emphasises anew that what is commonly regarded as the vacuum condensate is truly a property of the pion: it is simultaneously the chiral limit value of the in-pion condensate and proportional to the value of the chiral-limit pion's scalar form factor at zero momentum transfer.

Given Eq.\,(\ref{gmorR}), Eq.\,(\ref{gmor2}) is plainly a particular case of a more general statement; viz.,
\begin{eqnarray}
\label{chiPqQ}
{\cal S}^\zeta_{P_{qQ}}&:=&-\langle P_{qQ}|\mbox{\small $\frac{1}{2}$} (\bar q q+\bar Q Q) | P_{qQ} \rangle
=\frac{\partial}{\partial m_{qQ}^\zeta}m^2_{P_{qQ}} \rule{1em}{0ex}\\
%
&=& 
\frac{ \kappa^\zeta_{P_{qQ}} }{f^2_{P_{qQ}}} + m_{qQ}^\zeta
\frac{\partial}{\partial m_{qQ}^\zeta}  \left[ \frac{ \kappa^\zeta_{P_{qQ}} }{f^2_{P_{qQ}}}\right],
\end{eqnarray}
where $P_{qQ}$ is any pseudoscalar meson constituted from the current-quarks $q$, $Q$. 
The left-hand-side is this meson's scalar form factor at $Q^2=0$, which is here shown to be completely determined by the meson's leptonic decay constant and in-meson condensate, and their evolution with current-quark mass.

It is noteworthy that for each quark line within the bound-state, the $Q^2=0$ operator insertion in Eq.\,(\ref{chiPqQ}) acts as a differentiation of the affected dressed-quark propagator with respect to the current-quark mass.  On a dressed-quark in isolation, this would produce the vacuum chiral susceptibility \cite{Chang:2008ec} but here the observation establishes a clear connection between ${\cal S}$ and measurement of the chiral susceptibility within the hadron.

We have already considered the chiral-limit behaviour of ${\cal S}^\zeta_{P_{qQ}}$; viz., Eq.\,(\ref{gmor3}).  An exact result is also obtained in the heavy-quark limit: $m_Q\to \infty$, $m_q/m_Q \to 0$.  Following Ref.\,\cite{Ivanov:1998ms} one may demonstrate 
\begin{equation}
\label{mqmQlimit}
\kappa^\zeta_{P_{qQ}} \stackrel{m_Q\to\infty}{=} {\cal C}_P^\zeta\,,\;
f^2_{P_{qQ}} \stackrel{m_Q\to\infty}{=} \frac{\kappa^\zeta_{P_{qQ}}}{m_Q^\zeta}\,,\;
\end{equation}
where ${\cal C}_P^\zeta$ is an interaction-dependent constant.  Hence $m_{P_{qQ}} = m_q+m_Q$ and
\begin{equation}
\label{RPqQ}
{\cal S}^\zeta_{P_{qQ}}
\stackrel{m_Q\to\infty}{=} 2 \frac{ \kappa^\zeta_{P_{qQ}} }{f^2_{P_{qQ}}} =: 2 B_{P_{qQ}}^\zeta .
\end{equation}
It is notable that whilst for light current-quark masses, $f_P$ is an order parameter for DCSB, its evolution and essence are very different in the heavy-quark limit.

A single case remains; namely, pseudoscalar mesons constituted from current-quarks $Q_1$ and $Q_2$, with roughly equal masses, both of which become large: $m_{Q_1} \approx m_{Q_2}$, $m_{Q_2}\to \infty$.  Equations~(\ref{mqmQlimit}) are not valid in this instance.  Instead, the results depend on the nature of the interaction at short distances.  However, that is known to be Coulomb-like in QCD, so that one can show \cite{Bhagwat:2006xi}
\begin{eqnarray}
\label{mqmQlimit1}
\kappa^\zeta_{P_{Q_1Q_2}} &\stackrel{m_{Q}\to\infty}{=}& {\cal C}^\zeta_{P_Q}\,(M_{Q_1}+M_{Q_2})^3,\\
\label{mqmQlimit2}
f^2_{P_{Q_1Q_2}} &\stackrel{m_Q\to\infty}{=} & \frac{\kappa^\zeta_{P_{Q_1Q_2}}}{M_{Q_1}+M_{Q_2}},
\end{eqnarray}
with $M_Q^p:=M(-\mbox{\small $\frac{1}{4}$}m_{P_{Q_1Q_2}}^2)$, where $M(k^2)$ is the renormalisation-point-independent dressed-quark mass-function described, e.g., in Ref.\,\cite{Chang:2011vu}.  (In the limit considered here, $M_Q^p$ becomes equivalent to the ``pole-mass'' in the effective field theory for quarkonium systems.)  It follows therefore that, in precise analogy with Eq.\,(\ref{RPqQ}),
\begin{equation}
{\cal S}^\zeta_{P_{Q_1Q_2}}
\stackrel{m_{Q_1}\sim m_{Q_2}}{\stackrel{m_{Q_2}\to\infty}{=}}
2 \frac{ \kappa^\zeta_{P_{Q_1Q_2}} }{f^2_{P_{Q_1Q_2}}} =: 2 B_{P_{Q_1Q_2}}^\zeta \,.
\end{equation}

\begin{figure}[t]

\centerline{\includegraphics[clip,width=0.40\textwidth]{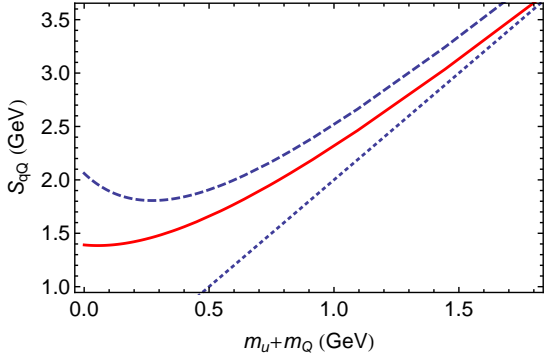}}

\caption{\label{Fig1}
Solid curve, ${\cal S}_{P_{qQ}}$; and dashed curve, ${\cal S}_{S_{qQ}}$.  The dotted line is the heavy-quark limit: $m_Q\to\infty$, $m_u/m_Q \to 0$ $\Rightarrow {\cal S}_{{qQ}} = m_u+m_Q$.  (${\cal S}^0_{P_{qQ}}=1.39\,$GeV, ${\cal S}^0_{S_{qQ}}=2.06\,$GeV; $m_u$ is fixed at 7\,MeV and $m_Q\geq m_u$.)}
\end{figure}

In order to explain and illustrate the nature of ${\cal S}^\zeta_{P_{qQ}}$, we have computed it using the symmetry-preserving regularisation and rainbow-ladder truncation of a vector$\,\times\,$vector contact-interaction that is described in Ref.\,\cite{Roberts:2011wy}.  The result, obtained with the light-quark parameters fixed therein, is depicted in Fig.\,\ref{Fig1}.  The behaviour is typical: ${\cal S}^\zeta_{P_{qQ}}$
is a positive-definite, monotonic function, bounded below by its chiral limit value ($B_{P_{qQ}}^{\zeta 0}$) and above by its large current-quark mass value ($2 B_{P_{qQ}}^{\zeta }$).

\begin{figure}[t]

\leftline{\includegraphics[clip,width=0.22\textwidth,height=0.30\textwidth]{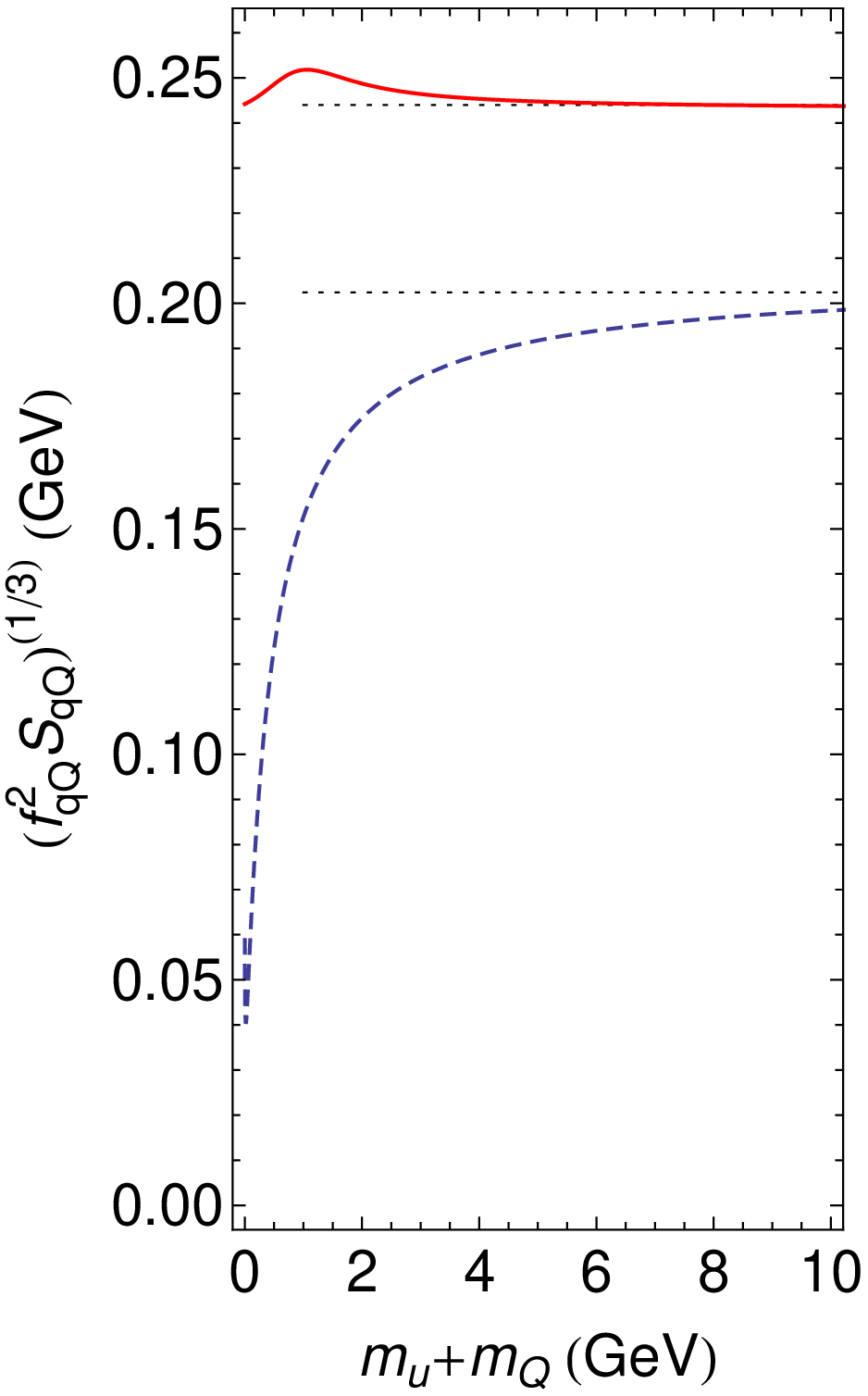}}
\vspace*{-40ex}

\rightline{\includegraphics[clip,width=0.225\textwidth,height=0.30\textwidth]{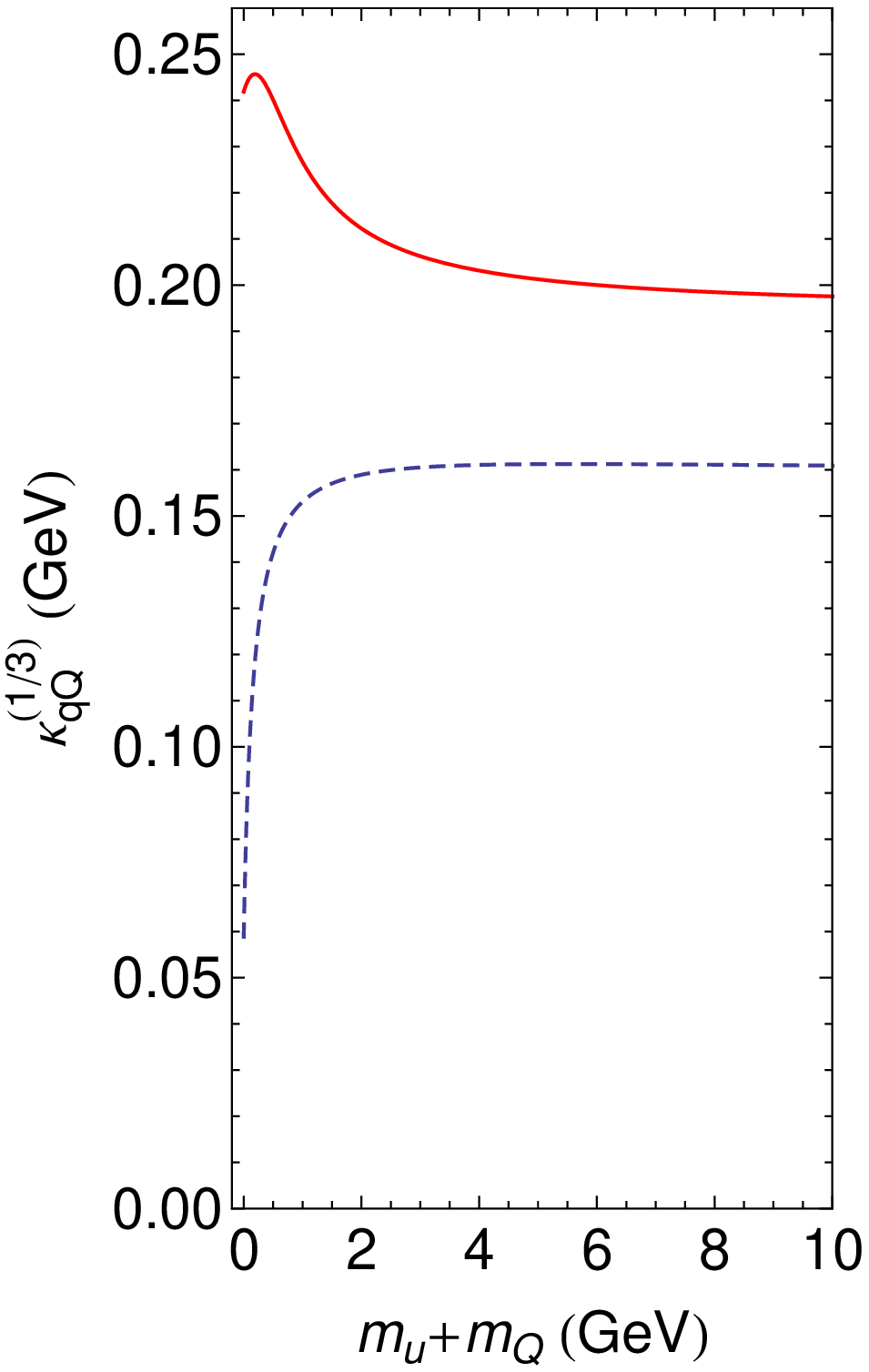}}

\caption{\label{Fig2}
\emph{Left panel} -- solid curve, $[f^2_{P_{qQ}} {\cal S}_{P_{qQ}}]^{1/3}$; dashed curve, $[f^2_{S_{qQ}} {\cal S}_{S_{qQ}}]^{1/3}$; and dotted lines, heavy-quark-limit values of $[2 \kappa_{qQ}]^{1/3}$ computed directly from Eqs.\,(\ref{fpigen}),(\ref{rhogen}) and Eqs.\,(\ref{fsigmagen}), (\ref{rhosigmagen}), respectively.
\emph{Right panel} -- $[\kappa_{qQ}]^{1/3}$ computed directly from Eqs.\,(\ref{fpigen}),(\ref{rhogen}) (pseudoscalar, solid curve) and Eqs.\,(\ref{fsigmagen}), (\ref{rhosigmagen}) (scalar, dashed curve).
The figure illustrates that $f^2_{{qQ}} {\cal S}_{{qQ}}$ is a smoothly varying measure of DCSB and confirms Eqs.\,(\protect\ref{F2eq1}), (\protect\ref{F2eq2}).
(NB.\ $\kappa^0_{P_{qQ}} = (0.24\,$GeV$)^3$; $m_u$ is fixed at 7\,MeV and $m_Q\geq m_u$.)
}
\end{figure}

With ${\cal S}^\zeta_{P}$, therefore, we have identified a quantity, defined for any and all pseudoscalar mesons, which directly measures the strength of helicity-coupling interactions within the hadron and whose value is between one- and two-times that strength.  Moreover,
\begin{equation}
(f_{P_{qQ}}^0)^2 {\cal S}^{\zeta 0}_{P_{qQ}} = \kappa^{\zeta 0}_{P_{qQ}}\;\mbox{and}\;
f_{P_{qQ}}^2 {\cal S}^\zeta_{P_{qQ}}
\stackrel{\mbox{\footnotesize heavy}}{\stackrel{\mbox{\footnotesize quark(s)}}{=}}
2 \kappa^{\zeta }_{P_{qQ}}\,,
\label{F2eq1}
\end{equation}
as illustrated in Fig.\,\ref{Fig2}, where $\kappa^{\zeta }_{P}$ is the  in-pseudoscalar-meson condensate introduced in Ref.\,\cite{Maris:1997tm}.  The matrix element ${\cal S}^\zeta_{P}$ thus appears ideal for use in extending the definition of in-hadron quark condensates to other states.

Further support for expansion of the in-hadron concept via this matrix element is provided by considering scalar mesons.  Applying the method of Ref.\,\cite{Maris:1997hd} to the vector Ward-Takahashi identity, we have established that
\begin{equation}
f_{S_{qQ}} m_{S_{qQ}}^2 = - \check{m}_{qQ} \rho_{S_{qQ}}^\zeta,
\label{mSqQ}
\end{equation}
where $\check{m}_{qQ}= m_q - m_Q$ and
\begin{eqnarray}
f_{S_{qQ}} K_\mu
& = & Z_2\, {\rm tr}_{\rm CD}\!\!\!
\int_k^\Lambda i\gamma_\mu S_q(k_+) \Gamma_{S_{qQ}}(k;K) S_Q(k_-)\,, \rule{2em}{0ex}
\label{fsigmagen}
\\
\rho^\zeta_{S_{qQ}}
& = & - Z_4\, {\rm tr}_{\rm CD}\!\!\!
\int_k^\Lambda S_q(k_+) \Gamma_{S_{qQ}}(k;K) S_Q(k_-) . \rule{2em}{0ex}\label{rhosigmagen}
\end{eqnarray}
The scalar meson leptonic decay constant changes sign under charge conjugation and vanishes for equal-mass constituents \cite{Maris:2000ig}.  Hence, Eq.\,(\ref{mSqQ}) does not reveal much about scalar meson masses in the chiral limit\footnote{The structure of light-quark scalar mesons is a contentious issue \protect\cite{RuizdeElvira:2010cs}.  Nevertheless, our result applies to any scalar meson that can be produced via $e^+ e^-$ annihilation.  It is not of experimental significance, however, if the pole is deep in the complex plane.} nor those composed of equal-mass heavy constituents.  On the other hand, much can be learnt in the heavy-quark limit. 
Indeed, one can prove analogues of Eq.\,(\ref{mqmQlimit}); viz.,
\begin{equation}
\label{SmqmQlimit}
\kappa^\zeta_{S_{qQ}} \stackrel{m_Q\to\infty}{=} {\cal C}_S^\zeta\,,\;
f_{S_{qQ}} \stackrel{m_Q\to\infty}{=} \frac{\kappa^\zeta_{S_{qQ}}}{m_Q^\zeta}\,,
\end{equation}
and hence
\begin{equation}
\label{HQRSP}
{\cal S}_{S_{qQ}} \stackrel{m_Q\to\infty}{=} 2 B_{S_{qQ}}^\zeta \stackrel{m_Q\to\infty}{=} 2 B_{P_{qQ}}^\zeta \stackrel{m_Q\to\infty}{=} {\cal S}_{P_{qQ}}\,,
\end{equation}
where ${\cal S}_{S_{qQ}}$ is defined by obvious analogy with Eq.\,(\ref{chiPqQ}).

We have also computed ${\cal S}^\zeta_{S_{qQ}}$ using the symmetry-preserving treatment of the contact-interaction \cite{Roberts:2011wy}.  Our result is depicted in Fig.\,\ref{Fig1}.  The behaviour is again typical; namely, ${\cal S}^\zeta_{S_{qQ}}$ is a positive-definite function that exceeds ${\cal S}^\zeta_{P_{qQ}}$ for all finite $m_{qQ}$ and approaches its heavy-quark limit from above.  Figure~\ref{Fig2} confirms the model-independent prediction in Eq.\,(\ref{HQRSP}); viz.,
\begin{equation}
f_{S_{qQ}}^2 {\cal S}^\zeta_{S_{qQ}} \stackrel{m_Q \to \infty}{=} 2 \kappa^{\zeta }_{S_{qQ}}\,.
\label{F2eq2}
\end{equation}
Quantitatively, the chiral-limit value of ${\cal S}^\zeta_{S_{qQ}}$ is interaction-dependent.  Within the framework of Ref.\,\cite{Roberts:2011wy}, the result is ${\cal S}^0_{S_{qQ}}= 4 M^0 (dM/dm)^0=2.06\,$GeV, where $M^0=0.36\,$GeV is the model's chiral-limit dressed-quark mass.  On the other hand, the qualitative connection to the dressed-quark mass, a bona-fide order parameter for DCSB which determines the so-called vacuum quark condensate, is model-independent.

We have demonstrated unique, model-independent relationships between ${\cal S}^\zeta_{P,S}$ and the in-hadron condensates that appear in mass formulae for pseudoscalar and scalar mesons.  Whilst such formulae do not exist for other mesons, the strength of the connections we've exhibited argues for the identification of an in-hadron condensate for each meson, $M$, with the product
\begin{eqnarray}
\label{chiMzeta}
\chi_M^\zeta &:=& {\cal S}^\zeta_{M} f_M^2,\\
{\cal S}^\zeta_{M}&:=&-\langle M|\mbox{\small $\frac{1}{2}$} (\bar q q+\bar Q Q) | M \rangle
=\frac{\partial}{\partial m_{qQ}^\zeta} m^2_{M},
\end{eqnarray}
where $m_M$ is the meson's mass and $f_M$, its leptonic decay constant.  The scalar case shows that a meaningful scale is determined even for systems with small $f_M$.

Within the framework of Ref.\,\cite{Roberts:2011wy}, one can readily evaluate results that follow for ground-state vector and axial-vector mesons; viz. (in GeV),
\begin{equation}
\begin{array}{cccccc}
{\cal S}_{\rho} & f_\rho & \chi_\rho^{1/3} & 
{\cal S}_{a_1} & f_{a_1} & \chi_{a_1}^{1/3} \\ 
1.33 & 0.129 & 0.281 & 2.30 & 0.089 & 0.263
\end{array}.
\end{equation}
A comparison with Figs.\,\ref{Fig1} and \ref{Fig2} makes evident a similarity between the: vector and pseudoscalar channels; and axial-vector and scalar channels.  This persists all the way to the heavy-quark limit whereat, owing to the suppression of hyperfine interactions, pseudoscalar and vector mesons are indistinguishable, as are scalar and axial-vector mesons, so that
\begin{equation}
f_{{V,A}_{qQ}}^2 \stackrel{m_Q\to\infty}{=} \frac{\kappa_{{V,A}_{qQ}}^\zeta}{m_Q^\zeta}\,,\;
\kappa_{{V,A}_{qQ}}^\zeta \stackrel{m_Q\to\infty}{=} \kappa_{{P,S}_{qQ}}^\zeta.
\end{equation}
%
The case of heavy-heavy $J=1$ states can also be argued by analogy with the $J=0$ states.

Baryons present a qualitatively different situation.  Owing to baryon-number conservation, there are no analogues of the meson decay constants in, e.g., Eqs.\,(\ref{fpigen}), (\ref{rhogen}), and hence no correspondents of the meson mass formulae.  Nonetheless, each baryon has a scalar form factor whose value at $Q^2=0$ is a perfect parallel to ${\cal S}_M$; viz.,
\begin{equation}
\label{SBaryon}
{\cal S}_{B_{1 2 3}}^\zeta:= -\langle B_{123} | \mbox{\small $\frac{1}{3}$} (\bar q_1q_1+\bar q_2q_2+\bar q_3q_3)|B_{123}\rangle,
\end{equation}
where $B_{123}$ is a baryon constituted from valence-quarks: $q_1$, $q_2$, $q_3$.  For baryons, too, ${\cal S}$ is a direct measure of the strength of helicity-coupling interactions within the hadron.  This commonality is a strength of our concept.

In the absence of decay constants, one can still identify a DCSB order parameter; viz., the baryon's mass itself.  This is clear once one appreciates that the nucleon's mass is approximately 1\,GeV because it is composed of three dressed-quarks, each of which has a mass $M\sim 350$\,MeV that owes primarily to DCSB \cite{Chang:2011vu}.  ${\cal S}_B$ is thus a dimensionless in-baryon chiral susceptibility: it measures the response to changes in the current-quark mass of a chiral order parameter which is intrinsic to the baryon.

\begin{figure}[t]

\centerline{\includegraphics[clip,width=0.42\textwidth]{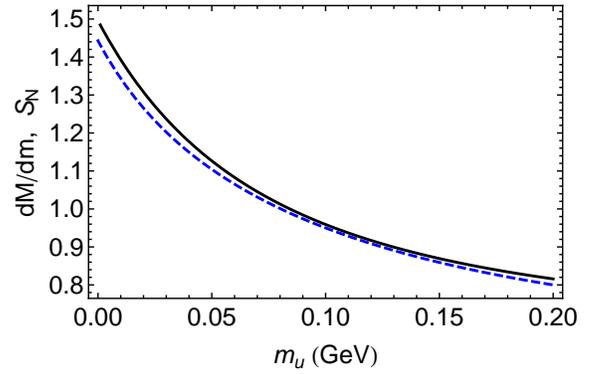}}

\caption{\label{Fig3}
Solid curve, ${\cal S}_{N}$: Eq.\,(\protect\ref{SBaryon}) for the nucleon; and dashed curve, $dM/dm$, where $M$ is the dressed-quark mass.  Both results computed using a symmetry-preserving regularisation of a vector$\,\times\,$vector contact interaction \protect\cite{Roberts:2011wy,Roberts:2011cf}: at $m=7\,$MeV, ${\cal S}_{N}=1.42$.}
\end{figure}

Using the framework of Ref.\,\cite{Roberts:2011wy}, the masses of the nucleon and $\Delta$-resonance, and their evolution with current-quark mass have been computed \cite{Roberts:2011cf}, with the result: for $0<m_\pi^2<0.5\,$GeV$^2$, $m_N \approx 1.03 \times (3 M)$, as a consequence of cancellation between complex binding effects.  It should therefore follow that ${\cal S}_N \approx dM/dm$ on this domain; viz., that helicity-coupling within the nucleon is as strong as that within ground state mesons.

This expectation is verified in Fig.\,\ref{Fig3}.  The quantitative results are interaction dependent.  Qualitatively, however, the comparison illustrates and highlights the capacity of ${\cal S}_B$ to serve as a gauge of DCSB within an internally consistent approach: in any theory the contrasting of ${\cal S}_B$ with an analogue of $dM/dm$ will provide a representative measure of the strength of DCSB within the baryon under consideration.

The last step is to identify a parallel for baryons of $\chi^\zeta_M$ in Eq.\,(\ref{chiMzeta}).  This appears problematic because, owing to baryon-number conservation, there is no baryonic analogue of $f_M$.  On the other hand, in contrast to ${\cal S}_M$, ${\cal S}_B$ is dimensionless and ${\cal S}_B\to 1$ in the heavy-quark limit.  Another inspection of the meson case provides an answer.

A homogeneous Bethe-Salpeter equation does not fix the normalisation of meson Bethe-Salpeter amplitudes.  An auxiliary condition must be implemented: one requires that an integral involving the amplitude and its conjugate must evaluate to some predetermined number, $N_M^2$.  The canonical normalisation condition constrains  the bound-state to produce a pole with unit reside in the quark-antiquark scattering matrix.  This may be represented as requiring $N_M^2=1$ (dimensionless).  One can naturally choose a different convention; e.g, consider the chiral-limit pion and rescale all elements in Eq.\,(\ref{genGpi}) so that $E_\pi(k;0)=B(k^2)$, where the latter function is the scalar piece of the dressed-quark self-energy in the chiral limit.  When evaluated now, the normalisation integral evaluates to $(N_\pi^0)^2 = (f_\pi^0)^2$, as a consequence of the axial-vector Ward-Takahashi identity \cite{Maris:1998hc}.  Although equality is not maintained away from the chiral limit, $N_\pi$, defined as described, is an order parameter for DCSB and vanishes in the heavy-quark limit.  Therefore, $N_{P_{qQ}}$ can mathematically be used to replace $f_{P_{qQ}}$ in Eq.\,(\ref{chiMzeta}).

The effect of this is readily illustrated within the framework of Ref.\,\cite{Roberts:2011wy}.  Normalising via $E_{P_{qQ}} = 2 \mu_{qQ}$, where $1/\mu_{qQ} = 1/m_q+1/m_Q$, one finds algebraically that $\forall \mu_{qQ}$,
\begin{equation}
N_{P_{qQ}} \rho_{P_{qQ}} = \tilde \chi_{P_{qQ}}
=  \mbox{\small $\frac{9}{2}$} \mu_{qQ} m_G^2
= N_{S_{qQ}} \rho_{S_{qQ}},
\end{equation}
which grows quickly from a chiral-limit value of $(0.243\,{\rm GeV})^3$ to $(0.307\,{\rm GeV})^3$ in the heavy-quark limit.  (NB.\ $m_G=0.132\,$GeV, fixed in the wide-ranging study of Ref.\,\cite{Roberts:2011wy}.)  It follows that $\tilde \chi_M^\zeta$, defined through the mass-normalised Bethe-Salpeter amplitude and $N_{P_{qQ}}$ computed therefrom, produces in-meson quark condensate mass-scales that are recognisably characteristic of DCSB.

Similar reasoning can be applied to the Faddeev equation.  In this case: the normalisation integral is connected with the value of the proton's Dirac form factor at $Q^2=0$; a mass-normalised baryon Faddeev amplitude produces a normalisation constant $N_B^2$ with dimensions of energy-cubed; and we have
\begin{equation}
\tilde\chi_{B_{123}}^\zeta :=  N_{B_{123}}^2\, {\cal S}_{B_{123}}^\zeta.
\end{equation}
To illustrate, we report that within the framework of Refs.\,\cite{Roberts:2011wy,Roberts:2011cf}, $N_N^2 = 3.40 M^3$ so that, using the value of ${\cal S}_{N}$ in Fig.\,\ref{Fig3}, $\tilde\chi_{N}=(0.623\,{\rm GeV})^3$.

The first rigorous demonstration that confinement restricts quark condensates to the interior of hadrons was made in connection with pseudoscalar mesons.  The in-pseudoscalar-meson condensate is a quantity with an exact expression in QCD.  We have proved that it can equally be represented through the pseudoscalar-meson's scalar form factor at zero momentum transfer, $Q^2=0$.  Subsequently, with the aid of a mass formula for scalar mesons, revealed herein, we showed that the in-scalar-meson condensate can be represented in precisely the same way.  By analogy, and with appeal to demonstrable results of heavy-quark symmetry, we argued that the $Q^2=0$ values of vector- and pseudovector-meson scalar form factors also determine the in-hadron condensates in these cases.  We also demonstrated that this expression for the concept of in-hadron quark condensates is readily extended to the case of baryons.
We therefore contend that via the $Q^2=0$ value of any hadron's scalar form factor, one can readily extract the value for a quark condensate in that hadron which is a reasonable and realistic measure of dynamical chiral symmetry breaking.

%
We acknowledge valuable input from A.~Bashir, S.\,J.~Brodsky, R.~Shrock and D.~J.~Wilson.
This work was supported by:
U.\,S.\ Department of Energy, Office of Nuclear Physics, contract no.~DE-AC02-06CH11357;
and U.\,S.\ National Science Foundation grant
no.\ NSF-PHY-0903991, part of which constitutes USA-Mexico collaboration funding in partnership with the Mexican agency CONACyT.


\begin{thebibliography}{99}

\bibitem{Nambu:1961tp}
  Y.~Nambu and G.~Jona-Lasinio,
  Phys.\ Rev.\  {\bf 122}, 345 (1961).

\bibitem{Casher:1974xd}
  A.~Casher and L.~Susskind,
  Phys.\ Rev.\  D {\bf 9}, 436 (1974).

\bibitem{Shifman:1978bx}
  M.~A.~Shifman, A.~I.~Vainshtein and V.~I.~Zakharov,
  Nucl.\ Phys.\  {\bf B147}, 385-447 (1979).

\bibitem{Turner:2001yu}
  M.~S.~Turner,
  astro-ph/0108103.

\bibitem{Brodsky:2009zd}
  S.~J.~Brodsky and R.~Shrock,
  Proc.\ Nat.\ Acad.\ Sci.\  {\bf 108}, 45 (2011).

\bibitem{Brodsky:2008be}
  S.~J.~Brodsky and R.~Shrock,
  Phys.\ Lett.\  B {\bf 666}, 95 (2008).

\bibitem{Brodsky:2010xf}
  S.~J.~Brodsky, C.~D.~Roberts, R.~Shrock and P.~C.~Tandy,
  Phys.\ Rev.\  C {\bf 82}, 022201(R) (2010).

\bibitem{Glazek:2011vg}
  S.~D.~Glazek,
  arXiv:1106.6100 [hep-th].

\bibitem{GellMann:1968rz}
  M.~Gell-Mann, R.~J.~Oakes and B.~Renner,
  Phys.\ Rev.\  {\bf 175}, 2195 (1968).

\bibitem{Flambaum:2005kc} V.~V.~Flambaum \textit{et al}.,
  Few Body Syst.\ \textbf{38} 31 (2006).

\bibitem{Maris:1997hd}
  P.~Maris, C.~D.~Roberts and P.~C.~Tandy,
  Phys.\ Lett.\  B {\bf 420}, 267 (1998).

\bibitem{Maris:1997tm}
  P.~Maris and C.~D.~Roberts,
  Phys.\ Rev.\  C {\bf 56}, 3369 (1997).

\bibitem{Langfeld:2003ye}
  K.~Langfeld \emph{et al}., 
  Phys.\ Rev.\  C {\bf 67}, 065206 (2003).

\bibitem{Lane:1974he}
  K.~D.~Lane,
  Phys.\ Rev.\  D {\bf 10}, 2605 (1974);
%
  H.~D.~Politzer,
  Nucl.\ Phys.\  B {\bf 117}, 397 (1976).

\bibitem{Banks:1979yr}
  T.~Banks and A.~Casher,
  Nucl.\ Phys.\  B {\bf 169}, 103 (1980).

\bibitem{Chang:2008ec}
  L.~Chang \emph{et al}., 
  Phys.\ Rev.\  C {\bf 79}, 035209 (2009)

\bibitem{Ivanov:1998ms}
  M.~A.~Ivanov, Yu.~L.~Kalinovsky and C.~D.~Roberts,
  Phys.\ Rev.\  D {\bf 60}, 034018 (1999).

\bibitem{Bhagwat:2006xi}
  M.~S.~Bhagwat, A.~Krassnigg, P.~Maris and C.~D.~Roberts,
  Eur.\ Phys.\ J.\  A {\bf 31}, 630 (2007).

\bibitem{Chang:2011vu}
  L.~Chang, C.~D.~Roberts and P.~C.~Tandy,
  arXiv:1107.4003 [nucl-th].

\bibitem{Roberts:2011wy}
  H.~L.~L.~Roberts \emph{et al}., 
  Phys.\ Rev.\  C {\bf 83}, 065206 (2011).

\bibitem{Maris:2000ig}
  P.~Maris, C.~D.~Roberts, S.~M.~Schmidt and P.~C.~Tandy,
  Phys.\ Rev.\  C {\bf 63}, 025202 (2001).

\bibitem{RuizdeElvira:2010cs}
  J.~Ruiz de Elvira, J.~R.~Pelaez, M.~R.~Pennington and D.~J.~Wilson,
  arXiv:1009.6204 [hep-ph].

\bibitem{Roberts:2011cf}
  H.\,L.\,L.~Roberts, L.~Chang, I.\,C.~Clo\"et and C.\,D.~Roberts,
  Few Body Syst.\  {\bf 51}, 1 (2011).

\bibitem{Maris:1998hc}
  P.~Maris and C.\,D.~Roberts,
  Phys.\,Rev.\,C\,{\bf 58}, 3659 (1998).

\end{thebibliography}
\end{document}